\documentclass[traditabstract]{aa}
\usepackage{graphicx,psfig}

\newcommand{\degree} {$^{\rm o}$}

\newcommand{\simless}{\mathbin{\lower 3pt\hbox
      {$\rlap{\raise 5pt\hbox{$\char'074$}}\mathchar"7218$}}} 
\newcommand{\simgreat}{\mathbin{\lower 3pt\hbox
     {$\rlap{\raise 5pt\hbox{$\char'076$}}\mathchar"7218$}}} 

\begin{document}

\newcommand{\hi}{\ion{H}{i}~}
\newcommand{\hii}{\ion{H}{ii}~}

   \title{The (sub-)millimeter SED of protoplanetary disks in the outskirts of the Orion Nebula Cluster}

   \author{L. Ricci         \inst{1}
\and          R. K. Mann          \inst{2} 
\and          L. Testi          \inst{1} 
\and          J. P. Williams          \inst{2} 
\and          A. Isella          \inst{3} 
\and          M. Robberto          \inst{4} 
\and A. Natta \inst{5} 
\and K. J. Brooks \inst{6}}


   \institute{  European Southern Observatory,
   Karl-Schwarzschild-Strasse 2, D-85748 Garching, Germany
            \and
                     Institute for Astronomy, University of Hawaii, 2680 Woodlawn Drive, Honolulu, HI 96822, USA
            \and
                     Division of Physics, Mathematics and Astronomy, California Institute of Technology, MC 249-17, Pasadena, CA 91125, USA                              
            \and
                     Space Telescope Science Institute, 3700 San Martin Dr., Baltimore, MD 21218, USA                             
            \and     
                     INAF - Osservatorio Astrofisico di Arcetri, Largo Fermi 5, I-50125 Firenze, Italy
			 \and 
			         Australia Telescope National Facility, P.O. Box 76, Epping, NSW 1710, Australia}

   \date{Received XXX 2010/ Accepted YYY ZZZZ}

   \titlerunning{Grain growth in Orion disks}

   \authorrunning{Ricci et al.}

\abstract{ We present the sub-mm/mm SED for a sample of eight young circumstellar disks in the outer regions of the Orion Nebula Cluster. New observations were carried out at 2.9~mm with the CARMA array and for one disk, 216-0939, at 3.3 and 6.8~mm with ATCA. By combining these new millimeter data with literature measurements at sub-millimeter wavelengths we investigate grain growth and measure the dust mass in protoplanetary disks in the Orion Nebula Cluster. These data provide evidence for dust grain growth to at least millimeter-sizes for the first time in a high-mass star forming region. The obtained range in sub-mm/mm spectral index, namely 1.5-3.2, indicates that for disks in the outskirts of the Orion Nebula Cluster (projected distance from the cluster center between about 0.4~pc and 1.5~pc) grain growth to mm sizes occurs in the same manner as disks in regions where only low-mass stars form. Finally, in our sample three disks are more massive than about $0.05\,M_\odot$, confirming that massive disks are present in the outer regions of the Orion Nebula.}



\maketitle


\section{Introduction}

Disks around young Pre-Main Sequence (PMS) stars are thought to be the cradles of planets, and for this reason they are often called ``proto-planetary disks''. Evidence for the very early stages of planet formation in protoplanetary disks has been obtained by a variety of different techniques, ranging from the optical to the millimeter-wave spectral domain. Compared to observations at shorter wavelengths the inspection of the sub-mm and mm emission of disks provides two key advantages for the investigation of the physical processes related to planet formation. First, except for the very inner regions of the disk, the dust optical depth at these long wavelengths is sufficiently low that the disk midplane can be probed. Second, since the maximum grain size to which a particular technique is sensitive to is of the order of the observational wavelength, the (sub-)millimeter spectral regime offers the chance to investigate grain growth up to mm/cm sizes (see Rodmann et al.~\cite{Rod06}).      

So far nearly all the information we have on the dust grain properties in protoplanetary disks has been derived from systems in low-mass Star Forming Regions (SFRs). Little information on the dust grains in young disks is present for high-mass SFRs. The analysis of dust in high-mass SFRs is particularly compelling since about 75\% of the observed Young Stellar Objects (YSOs) formed in a clustered environment, where high-mass stars form (Bressert et al.~\cite{Bre10}).  

Some evidence for dust grain growth in protoplanetary disks in the Orion Nebula Cluster (ONC) has been obtained at optical and infrared wavelengths. 
Throop et el.~(\cite{Thr01}) and Shuping et al.~(\cite{Shu03}) observed in optical and NIR the largest ONC silhouette disk (114-426). From the exinction curve of the background nebular emission they found that the NIR opacity of the disk is dominated by grains as large as about 10~$\mu$m, larger than the typical sub-micron sized grains found in the ISM (Mathis et al.~\cite{Mat77}). Using low-resolution spectroscopy in the mid-IR, Shuping et al.~(\cite{Shu06}) found evidence for grain growth to a few $\mu$m in the surface layers of eight disks in the Trapezium Cluster. 

During the last 15 years several ONC disks have been surveyed by interferometric observations at mm-wavelengths (e.g. Williams et al.~\cite{Wil05}, Eisner et al.~\cite{Eis08}, Mann \& Williams~\cite{Man09b}). However, the targeted disks were quite close to the main ionizing source of the cluster, the O6-type Trapezium star $\theta^1$ Ori C (typical projected distances lower than 0.2~pc by assuming a distance to the ONC of 400~pc, Menten et al.~\cite{Men07}, Sandstrom et al.~\cite{San07}, Kraus et al.~\cite{Kra09}). As a consequence, the high levels of free-free emission by ionized gas swamped the dust emission from the disks, making estimates of grain growth almost impossible to derive from multi-wavelength dust measurements in the millimeter spectral window.

\begin{figure}[tbp!]
\vspace{-1cm} 
\hspace{-1.2cm}
 \includegraphics[scale=0.55]{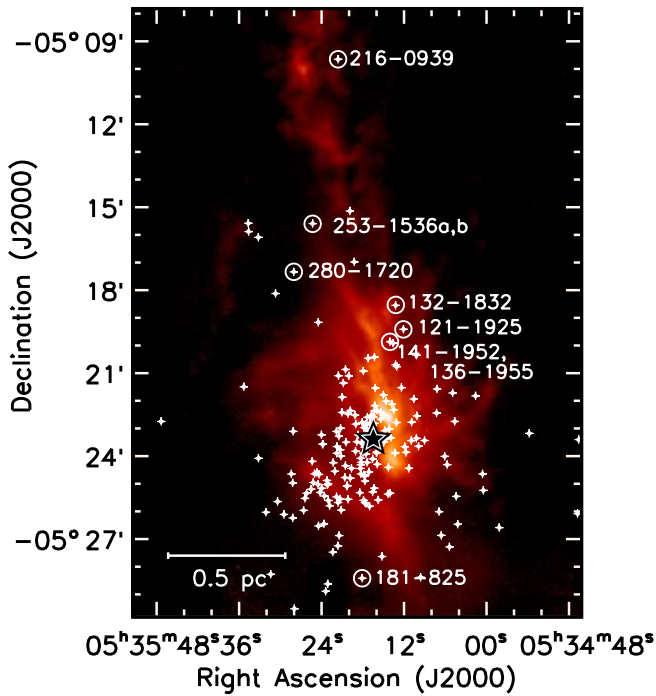}
 \vspace{-4.5cm}
 \caption{Location of the eight disks in our sample (within white circles) overlaid on 
a JCMT-SCUBA $850\,\mu$m image of the ONC (Johnstone \& Bally~\cite{Joh99}).  
The black and white star near the bottom of the image marks 
the position of the $\theta^1$ Ori C star.  
White crosses represent the positions of proplyds identified 
by HST observations. }
\label{fig:SCUBA_map} 
 \end{figure}

In this work we present new mm-wave interferometric observations for a sample of eight disk systems in the outskirts of the ONC. Because of their location relatively far from $\theta^1$ Ori C (projected distances between 0.4 and 1.5~pc, see Figure~\ref{fig:SCUBA_map}), information on grain growth can be derived from mm-wave observations. This allows us to probe the dust properties in these ONC disks over the entire sub-mm/mm SED.  

\section{Observations}
\label{sec:obs}

We describe here new mm-wave observations of eight young circumstellar disks in the ONC. The disks were selected for being detected within the SMA survey of protoplanetary disks in Orion (Mann \& Williams~2010, in press) with a 0.88~mm-flux density larger than about 15~mJy, and by being located in the outer regions of the ONC (i.e. projected distance from the Trapezium Cluster larger than 3$\arcmin$). This latter selection criterion was chosen to avoid contamination at mm-wavelengths from the non-uniform background emission and free-free emission from ionized gas typically observed in the inner ONC.
The disks in our sample are listed in Table~\ref{tab:obs}.

The dust thermal emission toward our sample of eight young disk systems in the 
ONC was observed with the Combined Array for 
Research in Millimeter Astronomy (CARMA) between 2010 Mar 26 and Apr 01. 
The array was in C configuration which provides baselines 
between 30 and 350 m. Observations were carried out at a central frequency of 102.5~GHz (2.92~mm), with a total bandwidth of 4~GHz.

\begin{figure}[tbp!]
 \includegraphics[scale=0.5]{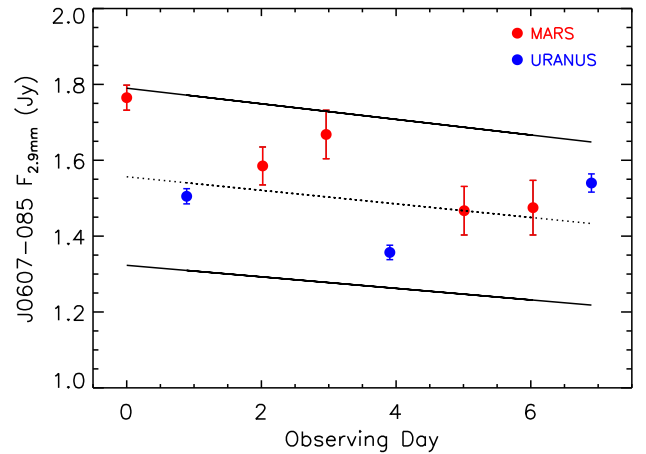}
 \caption{Flux density at 2.9~mm versus observing day for the QSO J0609-085 used as the phase and amplitude calibrator during the CARMA observations. Red circles represent fluxes obtained using Mars as the flux calibrator, whereas blue circles have been obtained using Uranus. The dotted line represents a linear fit of the data, and the two continuous lines refer to the $\pm 15\%$-levels from the linear fit.}
\label{fig:flux_calib} 
 \end{figure}

The raw visibilities for each night were calibrated and 
edited using the MIRIAD 
software package. Amplitude and phase 
calibration were performed through observations of the QSO J0607-085. 
Passband calibration was obtained by observing the QSO 0423-013. Mars 
and Uranus were used to set the absolute flux scale. The uncertainty on CARMA flux calibration is typically estimated to be $\sim$\,15\% and is due to uncertainties in the planetary models and in the correction for atmospheric effects and hardware instabilities. Figure~\ref{fig:flux_calib} shows the measured 2.9~mm-flux density of the QSO J0607-085 during the period covered by the CARMA observations. Both in the case that the QSO flux was slowly fading during the week of observations (with a light curve shown by the dotted line in the plot, which represents a linear fit of the data), and in the case that the QSO had a constant 2.9~mm flux density (equal to the weighted mean of the measured fluxes), the repeatability of our flux measurements is within the adopted $15\%$-level. $15\%$ is therefore a conservative estimate since part of the scatter may be due to intrinsic variability of the QSO on 1-day timescale. 

Maps of the dust continuum emission 
were obtained by standard Fourier inversion adopting natural 
weighting, and photometry was obtained in the image plane. The resulting FWHM of the synthesized beam is about 
2$\arcsec$.

One of our disks, 216-0939, was also observed with the Australia Telescope Compact Array (ATCA) and the new CABB digital filter bank, which provides a total continuum bandwidth of 2~GHz.
Observations were carried out at a central frequency of 91.000~GHz (3.294~mm) on 2009 Oct 13, and of 44.000~GHz (6.813~mm) on 2009 Oct 14. The ATCA array was in the hybrid H168 configuration, providing an angular resolution of about 3$\arcsec$ at 3.3~mm and 6.3$\arcsec$ at 6.8~mm. 

The gain was calibrated with frequent observations of 0529+075. The passband was calibrated using 1921-293, and the absolute flux scale was determined through observations of Uranus. The uncertainty on the calibrated flux is about 30\% at 3.3~mm and 20\% at 6.8~mm.
The MIRIAD package was used for visibilities calibration, Fourier inversion, deconvolution and imaging. 

\begin{table*}

\centering \caption{ Sub-Millimeter and Millimeter Flux Densities and Disk Sizes} \vskip 0.1cm
\begin{tabular}{lccccccccc}
\hline \hline
\vspace{-2mm}\\
Object  & $F_{\rm{0.88 mm}}$ & $F_{\rm{2.9 mm}}$ & $F_{\rm{3.3 mm}}$ & $F_{\rm{6.8 mm}}$ & $\alpha$ &   SMA Size    &      HST Size & $\kappa_{\rm{2.9 mm}}^{q=2.5}$ & $M_{\rm{disk}}^{q=2.5}$  \vspace{1mm} \\
        & (mJy)              & (mJy)             & (mJy)             & (mJy)             &          &     (arcsec)       &      (arcsec) & (cm$^2$g$^{-1}$) & ($M_{\odot}$)     \vspace{1mm} \\
(1)     &    (2)             &	(3)	             &     (4)           &       (5)         &   (6)    &     (7)            &     (8)     &  (9) & (10)  \\
\hline
\vspace{-3mm} \\
121-1925    & 15.0$\pm$1.5                 & 2.4$\pm$0.4 & ... & ... & 1.5$\pm$0.4  & Unresolved  & 0.8 & 0.003 & 0.1 \\
132-1832    & 16.5$\pm$1.7                 & $<$2.7        & ... & ... & $>$1.5     & 1.1$\pm$0.4 & 1.3 & ... & ... \\
136-1955    & 77.6$\pm$1.2                 & 1.8$\pm$0.4 & ... & ... & 3.2$\pm$0.5  & 0.5$\pm$0.1 & ... & 0.02 & 0.006 \\
141-1952    & 30.6$\pm$1.2                 & 1.9$\pm$0.4 & ... & ... & 2.3$\pm$0.5  & 0.4$\pm$0.1 & 0.7 & 0.02 & 0.003 \\
181-825     & 54.8$\pm$1.0                 & 4.4$\pm$0.5 & ... & ... & 2.1$\pm$0.4  & Unresolved  & 0.7 & 0.01 & 0.05 \\
216-0939    & 91.9$\pm$1.3                 & 3.5$\pm$0.6 & 2.5$\pm$0.3 & 0.34$\pm$0.05 & 2.7$\pm$0.2 & 1.6$\pm$0.1 & 2.6 & 0.04 & 0.02  \\
253-1536$^{\rm{a}}$   & 134.2$\pm$1.0  & 7.9$\pm$0.7   & ... & ... & 2.6$\pm$0.4 & 1.5$\pm$0.1 & 1.5 &  0.04 & 0.05 \\
280-1720    & 21.8$\pm$0.7                 & $<$1.8        & ... & ... & $>$2.1      & Unresolved  & 0.8 & ... & ... \\

\hline
\end{tabular}
\begin{flushleft}
1) Object name.
2) 0.88~mm-flux density (from Mann \& Williams~2010, in press).
3) 2.9~mm-flux density.
4) 3.3~mm-flux density.
5) 6.8~mm-flux density.
6) Sub-mm/mm spectral index derived by fitting with a power-law the data shown in the table. The errors account for the uncertainty on the flux scale at each waveband.
7) FWHM along the disk major axis from a gaussian fit in the SMA images.
8) Disk diameter from HST images (Smith et al.~\cite{Smi05}, Ricci et al.~\cite{Ric08}).
9) 2.9~mm-dust opacity derived by the SED-fitting procedure (and the adopted dust model) assuming $q=2.5$.
10) Disk mass derived by the SED-fitting procedure assuming $q=2.5$ (see Sect.~\ref{sec:mass}).
a) The SMA observations of 253-1536 show a weaker secondary component ($F_{\rm{0.88 mm}}=37.6\pm1.0$ mJy). The spectral index of this system has been calculated by using for $F_{\rm{0.88 mm}}$ the sum of the SMA fluxes separately measured for the two circumstellar disks. For this reason the derived spectral index and disk mass are descriptive of the whole system rather than of the two disks separately. 



\end{flushleft}

\label{tab:obs}

\end{table*}

\section{Results}
\label{sec:results}

\begin{figure}[tbp!]
\vspace{-0.3cm}
\hspace{-0.5cm} 
 \includegraphics[scale=0.55]{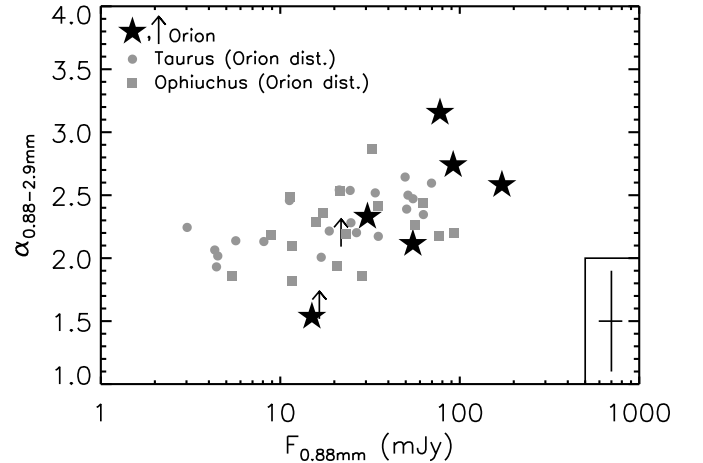}
\vspace{-0.3cm}
 \caption{ 
Spectral index between 0.88 and 2.9~mm plotted against the flux at 0.88~mm. Black stars and lower-limits are for the ONC disks, whereas grey circles and squares are for class II disks in Taurus (Ricci et al.~\cite{Ric10a}) and Ophiuchus (Ricci et al.~\cite{Ric10b}), respectively. The fluxes of the Taurus and Ophiuchus disks have been scaled to the ONC-distance adopted in this work (400~pc) to allow a comparison with the ONC disks. In the lower-right corner of the plot, the typical errorbars of the plotted quantities are shown. }
\label{fig:flux_alpha} 
 \end{figure}

Table~\ref{tab:obs} reports the sub-mm/mm flux densities obtained for our sample of eight young circumstellar disks in the outskirts of the ONC.
Column (6) lists the sub-mm/mm spectral index $\alpha$ ($F_{\nu} \sim \nu^{\alpha}$) as derived from the results of our observations.
Whereas all the eight disks in our sample have been detected with the SMA at 0.88~mm (Mann \& Williams~2010, in press), we detected six of them at 2.9~mm with CARMA. For the two undetected disks, 132-1832 and 280-1720, the upper-limits in the 2.9~mm-flux density translate into lower-limits for the sub-mm/mm spectral index. The range in sub-mm/mm spectral index spanned by our sample, i.e. $\sim$~1.5-3.2,  is very close to what observed for class II YSOs in low-mass SFRs (see Figure~\ref{fig:flux_alpha}). This indicates that the properties of dust grains for disks in the outskirts of the ONC are very similar to those found in disks in low-mass SFRs. Furthermore we did not find any clear trend between spectral index $\alpha$, HST size, disk mass, or distance from $\theta^1$ Ori C.

\subsection{Grain Growth}
\label{sec:grain_growth}

Information on the grain-size distribution of dust in the outer regions of protoplanetary disks can be obtained by analyzing the disk SED at sub-mm/mm wavelengths (e.g. Draine~\cite{Dra06}). More specifically, for a disk whose sub/mm-mm flux is entirely due to optically thin dust emission in the Rayleigh-Jeans regime, a very simple relation between the measurable spectral index $\alpha$ of the sub-mm/mm SED and the spectral index $\beta$ of the dust opacity\footnote{At long wavelengths the dust opacity law is well represented by a power law $\kappa_{\nu} \propto \nu^{\beta}$.} holds, namely $\beta = \alpha - 2$. 

Values of $\beta$ lower than the value found for the unprocessed ISM dust population ($\beta_{\rm{ISM}} \sim$ 1.7-2) can be obtained only with dust population where grains as large as at least $\sim$ 1~mm are present (see e.g. Natta et al.~\cite{Nat07}).
Deviations from the aforementioned relation between $\alpha$ and $\beta$ can be accounted for by using physical disk models to fit the disk SED together with the knowledge of the spatial extension of the disk, which is needed to estimate the impact of the inner optically thick regions on the long-wave spectrum (see Testi et al.~\cite{Tes01}). By doing this for 38 class II disks around low-mass PMS stars in the Taurus-Auriga and Ophiuchus SFRs, Ricci et al.~(\cite{Ric10a}, 2010b) found $\alpha - \beta = 1.8 \pm 0.1$. 
All disks in our sample have outer radii larger than $\sim$ 100~AU (Table~\ref{tab:obs}, col. 7 and 8), all the model fits have only a minimal contribution from optically thick emission (see discussion in Ricci et al.~\cite{Ric10a}).    
Applying the $\alpha$-$\beta$ relation to the Orion disks, we obtain values for the dust opacity spectral index $\beta$ ranging between about 0 for 121-1925
and 1.4 for 136-1955. Apart for the case of 136-1955 whose 1$\sigma$-interval of $\beta$ is consistent with $\beta_{\rm{ISM}}$, for all the other five disks detected with CARMA the constrained $\beta$-values are significantly lower than $\beta_{\rm{ISM}}$, and thus we found firm evidence of grain growth to $\sim$~mm-sized pebbles for the first time in the ONC, and, more generally, in the cluster environment of a high-mass SFR.

Note that, to derive the results described so far, we assumed that all the measured sub-mm/mm flux comes from dust emission only. At mm-wavelengths some contribution to the measured flux from ionized gas has been found in the literature. This contribution can be constrained from observations at longer cm-wavelengths, where free-free emission from ionized gas dominates over the dust emission. For disks in low-mass SFRs the presence of ionized gas is probably induced by the ionizating flux from the central star, and it has been found to contribute typically for 25\% or less at 7~mm and a few percent at most at 3~mm (Wilner et al.~\cite{Wiln05}, Rodmann et al.~\cite{Rod06}, Lommen et al.~\cite{Lom09}). For disks in the inner ONC at angular separations lower than about 1 arcmin from the main ionizing source of the entire region, $\theta^1$ Ori C (projected separations lower than about 0.1~pc), free-free emission has been typically found to dominate the emission at wavelengths close to 3~mm and longer (Williams et al.~\cite{Wil05}, Eisner et al.~\cite{Eis08}, Mann \& Williams~\cite{Man09b}). In order to get accurate estimates for dust grain growth from multi-wavelengths observations in the millimeter, it is important to consider disks which are further away from $\theta^1$ Ori C, for which the free-free emission does not swamp the mm-wave emission from dust. Our sample is made of sources which are at projected distances between about 0.4 and  1.5~pc. Since our disks are much further away in terms of projected distance to $\theta^1$ Ori C than the disks observed in the literature, in the rest of the paper we will assume that no contribution from free-free emission is present for the disks in our sample. This is supported by the fact that the derived spectral slope $\alpha$ for our sources do not correlate with the projected distance to $\theta^1$ Ori C, as one may expect if significant contamination from free-free emission was present in our data. Furthermore, in the case of 216-0939 for which new data at 3~mm and 7~mm have been obtained, the sub-mm/mm SED does not show any change of the spectral slope (Figure~\ref{fig:216-0939_SED}), as it would be expected if free-free emission was significant at these wavelenghts (see e.g. Williams et al.~\cite{Wil05}). However, future observations at longer cm-wavelengths are needed to really constrain the impact of free-free emission and test this assumption.

\subsection{The case of 216-0939}
\label{sec:216}

\begin{figure}[tbp!]
\vspace{-0.3cm}
 \includegraphics[scale=0.5]{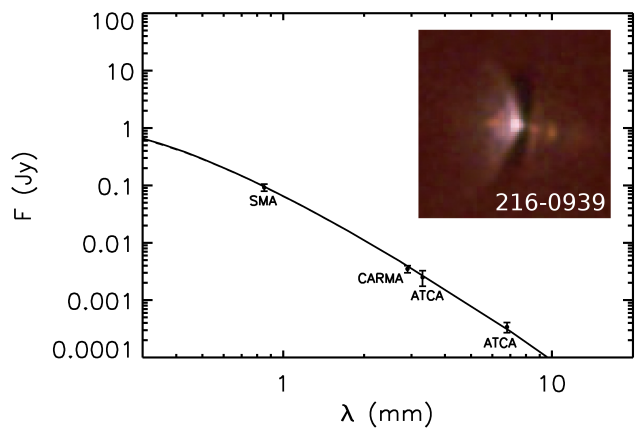}
\vspace{-0.3cm}
 \caption{Fit of the sub-mm/mm SED of the 216-0939 Orion disk with a two-layer passive flared disk model. The parameters of the model shown here are described in Sect.~\ref{sec:216}. The uncertainties in the data points account for the uncertainty on the flux-scale. The color image of the 216-0939 disk is from HST observations by Smith et al.~(\cite{Smi05}).}
\label{fig:216-0939_SED} 
 \end{figure}

Figure~\ref{fig:216-0939_SED} shows the fit of the sub-mm/mm SED of 216-0939, which is the best characterized disk in our sample, with two-layer (surface$+$midplane) models of flared disks heated by the radiation of the central star (Chiang \& Goldreich~\cite{Chi97}, Dullemond et al.~\cite{Dul01}). For the stellar parameters we considered a PMS star with a K5-spectral type (from Hillenbrand~\cite{Hil97}), which we converted into an effective temperature of 4350~K using the temperature scale of Schmidt-Kaler~(\cite{Sch82}). We then derived a stellar mass of about 1.2~$M_{\odot}$ and a stellar luminosity of 2.5~$L_{\odot}$ by placing the PMS star onto a 1 Myr-isochrone from the Palla \& Stahler~(\cite{Pal99}) PMS evolutionary models.
For the disk we adopted a truncated power-law mass surface density with an exponent of 1 (as found by Andrews \& Williams~\cite{And07} from disks in Taurus and Ophiuchus) and an outer radius of about 290~AU from the size of the mm-emission in the SMA map (Table~\ref{tab:obs}). A disk inclination of 75\degree\ has been taken from an analysis of optical HST images (Smith et al.~\cite{Smi05}). From the SED-fitting procedure (Fig.~\ref{fig:216-0939_SED}) we derived $\beta = 1.0 \pm 0.3$, which is indicative of grain growth in the 216-0939 disk. 

By using the same dust model discussed in Ricci et al.~(\cite{Ric10a}, 2010b), i.e. spherical porous grains made of silicates, carbonaceous materials and water ice (abundance of each species from Semenov et al.~\cite{Sem03}), we could get information on the grain-size distribution, here assumed to be a truncated power-law with index $q$ and maximum grain size $a_{\rm{max}}$.
A good fit of the SED could be obtained only for $q \simless 3.5$.
Grain-size distributions with $q$-values of 3 and 2.5 fit equally well the sub-mm/mm SED but with different $a_{\rm{max}}$-values, namely about 2.5 and 2~mm, respectively, but here it is important to note that the precise estimate for the maximum grain size $a_{\rm{max}}$ can strongly depend on the adopted dust model.


\subsection{Derived disk masses}
\label{sec:mass}

Other than providing information on the grain-size distribution, the SED-fitting procedure described in the last section returns estimates for the dust mass in the disk, and, by assuming the ISM-value of about 100 for the gas-to-dust mass ratio, for the total disk mass.
Column (9) in Table~\ref{tab:obs} reports the derived disk masses for the six disks detected at millimeter wavelengths, for which information on the dust properties could be inferred. For these estimates we considered a $q-$value of 2.5 because for the two disks showing the lowest mm-spectral indeces, 121-1925 and 181-825, models with $q \simgreat 3$ cannot reproduce the data. This shows that the grain size distribution in these disks is shallower than found for the ISM ($q_{\rm{ISM}} \approx 3.5$, Mathis et al.~\cite{Mat77}). This result, which, as described in Ricci et al.~(\cite{Ric10a}) does not depend on the adopted dust model, is another indicator for dust processing in these disks.      

The range in disk mass covered by our sample is 0.003$-$0.1~$M_{\odot}$. This shows that massive disks are present in the outer regions of the ONC, confirming recent results by Mann \& Williams (\cite{Man09a}, 2010, in press). Note that 121-1925 is the faintest disk at 0.88~mm but is nevertheless detected at 0.88~mm. We thus derive a very low value for $\alpha$. In our analysis, this implies a very low value for $\beta$, and a very significant inferred grain growth, suggesting a very low value for the dust opacity coefficient at 2.9~mm (see col. 9 in Table~\ref{tab:obs}), and consequently a large disk mass. 

\section{Summary}
\label{sec:summary}

We present new millimeter interferometric observations of eight disks in the outskirts of the ONC.
Together with sub-mm data in the literature, these new data allow us to study dust properties, like grain growth and disk mass, from the entire sub-mm/mm SED.
Except for one disk, 136-1955, whose emission is consistent within the errorbars with a ISM-like dust population, we found evidence, for the first time in cluster environment in a high-mass SFR, of grain growth to at least mm-sized pebbles for 5 out of the 6 detected disks at mm-wavelengths. As for the disk masses, estimated by using a dust model which accounts for grain growth in the calculation of the dust opacities, our results
confirm that massive disks ($M\geq 0.05\,M_\odot$) can be found in the outskirts of the ONC.

\begin{acknowledgements}
L. R. wishes to thank the support astronomers in Narrabri for their help during ATCA observations. L. R. aknowledges the PhD fellowship of the International Max-Planck-Research School. 
J. P. W. is supported by the NSF through grant AST08-08144. A. I. is supported by the Jet Propulsion Laboratory (JPL) funded by NASA through the Michelson Fellowship Program.
Support for CARMA construction was derived from the states of California, Illinois, and Maryland, the James S. McDonnell Foundation, the Gordon and Betty Moore Foundation, the Kenneth T. and Eileen L. Norris Foundation, the University of Chicago, the Associates of the California Institute of Technology, and the National Science Foundation. Ongoing CARMA development and operations are supported by the National Science Foundation under a cooperative agreement, and by the CARMA partner universities. The Australia Telescope Compact Array is part of the Australia Telescope which is funded by the Commonwealth of Australia for operation as a National Facility managed by CSIRO. 
\end{acknowledgements}



\end{document}